\documentclass[useAMS,usenatbib]{mn2e}

\title[Complex particle acceleration processes in
  3C\,105 and 3C\,445]{Complex particle acceleration processes in the
  hotspots of 3C\,105 and 3C\,445\thanks{Based on VLT programs 72B-0360B, 70B-0713B, 267B-5721, and HST program 10434.}}
\author[M. Orienti et al. ]
  {M. Orienti$^{1,2}$\thanks{E-mail: orienti@ira.inaf.it},
M.A. Prieto$^3$, G. Brunetti$^2$, K.-H. Mack$^2$, F. Massaro$^4$, D.E. Harris$^4$\\
$^{1}$Dipartimento di Astronomia, Universit\`a di Bologna, via Ranzani 1,
I-40127, Bologna, Italy \\
$^{2}$Istituto di Radioastronomia - INAF, Via P. Gobetti 101, I-40129 Bologna, Italy\\
$^3$Instituto de Astrof\'{i}sica de Canarias, c/ V\'{i}a L\'actea s/n,
E-38205  La Laguna (Tenerife), Spain\\
$^4$Smithsonian Astrophysical Observatory, 60 Garden Street, Cambridge, MA 02138, USA
}
\date{Received \today; accepted ?}

\pagerange{\pageref{firstpage}--\pageref{lastpage}} \pubyear{2002}

\def\LaTeX{L\kern-.36em\raise.3ex\hbox{a}\kern-.15em
    T\kern-.1667em\lower.7ex\hbox{E}\kern-.125emX}

\begin{document}

\label{firstpage}

\maketitle

\begin{abstract}

We investigate the nature of the broad-band emission associated with the
low-power radio hotspots 3C\,105 South and 3C\,445 South. Both hotspot
regions are resolved in multiple radio/optical components.
High-sensitivity radio VLA, NIR/optical VLT and HST, and X-ray {\it Chandra}
data have been used to construct the multi-band spectra of individual
hotspot components. The radio-to-optical spectra of both hotspot regions
are well fitted by a synchrotron model with steep spectral indices
$\sim 0.8$ and break frequencies between $10^{12}-10^{14}$ Hz.
3C\,105 South is resolved in two optical components: a primary one,
aligned with the jet direction and possibly marking the first
jet impact with the surrounding medium, and a
secondary, further out from the jet and extended in a direction
  perpendicular to it. This secondary region is interpreted as a 
splatter-spot formed by the deflection of relativistic plasma  
from the primary hotspot. 
Radio and optical images of 3C\,445 South show a spectacular 10-kpc arc-shape
structure characterized by two main components, 
and perpendicular to the jet direction.
HST images in I and B bands further resolve the brightest 
components into thin elongated features.
In both 3C\,105 South and 3C\,445 South the main hotspot components
are enshrouded by diffuse optical emission
on scale of several kpcs, indicating that very high energy
particles, possibly injected at strong shocks, are
continuously re-accelerated in situ by additional acceleration
mechanisms. We suggest that stochastic processes, linked to
turbulence and instabilities, could provide the required additional
re-acceleration.

\end{abstract}


\section{Introduction}

Radio hotspots are bright and compact regions located at the end of
powerful radio galaxies 
\citep[FRIIs,][]{fr2} and
considered to be the working surfaces of supersonic jets. In these
regions,
the jet emitted by the active galactic nucleus
(AGN) impacts on the surrounding ambient medium producing a shock that
may re-accelerate relativistic particles transported by the jet and
enhance the radio emission. 
Electrons responsible for synchrotron emission in the
optical band must be very energetic (Lorentz factor $\gamma >
10^{5}$),
and therefore with short radiative lifetime.
Consequently the detection of optical emission from hotspots
supports the scenario where the emitting electrons are accelerated
at the hotspots, possibly by strong shocks generated by the impact of 
the jet with the ambient medium \citep{meise89,meise97,gb03}. 
The detection of X-ray synchrotron counterparts of radio hotspots
  would imply the
  presence of electrons with even higher energies. 
  However the main radiation process
  responsible for the X-ray emission seems to differ between high and
  low luminosity hotspots \citep{hardcastle04}. 
  In bright hotspots, like Cygnus A and 3C\,295, the X-ray
  emission is produced by synchrotron-self Compton (SSC) in the presence
  of a magnetic field that is roughly in equipartition, while in 
low-luminosity hotspots, like 3C\,390.3, the emission at such high
  energies is likely due to
  synchrotron radiation \citep{hardcastle07}.\\
The discovery of optical emission extended to kpc scale 
questions the standard shock acceleration model, suggesting that 
other efficient
mechanisms must take place across the hotspot region.
Although it may seem an uncommon phenomenon due to the
difficulty to produce high-energy electrons on large scales, deep
optical images showed that diffuse
optical emission is present in a handful of hotspots: 3C\,33, 
3C\,111, 3C\,303, 3C\,351
\citep{valta99}, 3C\,390.3 \citep{aprieto97}, 
3C\,275.1 \citep{cheung05}, Pictor A \citep{thomson95}, 
and 3C\,445 \citep{aprieto02}. 
A possible mechanism able to keep up the optical emission in the post-shock
region on kpc scale is a continuous, relatively efficient, stochastic 
mechanism\footnote{More recently these stochastic
mechanisms have been also proposed for
the acceleration of ultra-high energy cosmic-rays in the lobes of
radiogalaxies \citep{hardcastle09}.}.\\
The sample of low-power hotspots presented by \citet{mack09} is
characterized by low magnetic field strengths between 40 and 130
$\mu$G, a factor 2
to 5 lower than that estimated in hotspots with optical counterparts
previously studied in the literature. A surprisingly high optical
detection rate ($\geq$ 45\%)
of the hotspots in this sample was found, and in most cases 
the optical counterpart extends on kpc scales. This
is the case of 3C\,445 South, 3C\, 445 North, 3C\,105 South and
3C\,227 West \citep{mack09}.\\
This
paper focuses on a multi-band, from radio to X-rays, 
high spatial resolution study of the two
most interesting cases among the low-luminosity hotspots from
  \citet{mack09}, 3C\,105 South and 3C\,445 South, in which the
hotspot regions are resolved into multiple components. 
3C\,105 is hosted by a narrow-line radio galaxy (NLRG) at redshift
  $z=0.089$ \citep{tadhunter93}. At this redshift 1$^{\prime\prime}$
  corresponds to 1.642 kpc. The radio source 3C\,105 is about
  330$^{\prime\prime}$ (542 kpc) in size, and the hotspot complex 3C\,105 South is
  located about 168$^{\prime\prime}$ (276 kpc) from the core in the
  south-east direction. 3C\,445 is hosted by a broad-line radio galaxy
  (BLRG) 
at redshift $z=0.05623$ \citep{eracleous94}. At this redshift
1$^{\prime\prime}$ corresponds to 1.077 kpc. The radio source 3C\,445
is about 562$^{\prime\prime}$ (608 kpc) in size, and the hotspot
complex 3C\,445 South is located 270$^{\prime\prime}$ (291 kpc)
south of the core.

Throughout this paper, we assume the following cosmology: $H_{0} =
71\; {\rm km/s\, Mpc^{-1}}$, 
$\Omega_{\rm M} = 0.27$ and $\Omega_{\rm \Lambda} = 0.73$,
in a flat Universe. The spectral index
is defined as 
$S {\rm (\nu)} \propto \nu^{- \alpha}$.\\

\section{Observations}

\subsection{Radio observations}

VLA observations at 1.4, 4.8, and 8.4 GHz 
of the radio hotspots 3C\,445 South and
3C\,105 South were carried out in July 2003 (project code AM772)
with the array in
A-configuration. Each source was observed for about half an
hour at each frequency, spread into a number of scans
interspersed with other source/calibrator scans in order to improve
the $uv$-coverage. About 4 minutes were spent on the
primary calibrator 3C\,286, while secondary phase calibrators
were observed for 1.5 min about every
5 min. Data at 1.4 and 4.8 GHz were previously published by \citet{mack09}.
The data reduction was carried out following the standard procedures
for the VLA implemented in the NRAO AIPS package.
Final images were produced after a few phase-only self-calibration
iterations. The r.m.s. noise level on the image plane is negligible if compared
to the uncertainty of the flux density due to amplitude calibration
errors that, in this case, are estimated to be $\sim$3\%.\\
Besides the {\it full-resolution} images, we also produced {\it
  low-resolution} images at both 4.8 and 8.4 GHz, 
using the same $uv$-range, image sampling and restoring beam of the
1.4 GHz data. These new images were obtained with natural grid 
weighting in order to mitigate the differences in the sampling density 
at short spacing, and to perform a robust spectral analysis. \\

\subsection{Optical observations}

For both 3C\,105 South and 3C\,445 South, VLT high spatial resolution
images in standard filters taken with both ISAAC in J-, H-, K-, and FORS
in I-, R-, B- and U- bands are used in this work. All the images have
excellent spatial resolutions in the range of 0.5$^{\prime\prime}$ $<$ FWHM $<$
0.7$^{\prime\prime}$. Details on the observations and data reduction are given in
\citet{mack09}. The pixel scale of the ISAAC images is 0.14 arcsec
pixel$^{-1}$. In the case of the FORS images the pixel scale is 0.2
arcsec pixel$^{-1}$, with the exception of the I-band where it is
  0.1 arcsec pixel$^{-1}$.\\ 
Further HST observations on 3C\,445 South only, were obtained with
the ACS/HRC camera on 7th July, 2005  
in the filters F814W (I-band, exposure time $\sim$ 1.5 hr) and
F475W (B-band, exposure time $\sim$ 2.3 hr).\\  
For science analysis we used the ``*drz'' images delivered 
by the HST ACS pipeline. These final images are calibrated, 
cosmic-ray cleaned, geometrically corrected, and drizzle-combined, 
provided in electrons per sec. The final pixel scale of the 
drizzled images is 0.025$^{\prime\prime}$$\times$0.025$^{\prime\prime}$ per pixel. 
The flux calibration was done using the standard HST/ACS procedure
that relies on the PHOTFLAM keyword in the respective image headers.
The quality of the pipeline-delivered images was adequate for the
purposes of analyzing the hotspot region.

\begin{table*}
\caption{Radio flux density and angular size of the hotspot
  components. \newline {\it Note 1}: deconvolved angular sizes from a Gaussian
  fit. \newline {\it Note 2}:
the angular sizes are derived from the lowest contour on the image
plane; \newline
{\it Note 3}: The diffuse emission is estimated by subtracting
the flux density of SW and SE from the total flux density (see Section
5.3).}
\begin{center}
\begin{tabular}{|c|r|r||r|r|r|r|r|r|}
\hline
Source&Comp.&z&scale&S$_{1.4}$&S$_{4.8}$&S$_{8.4}$&$\theta_{\rm maj}$&$\theta_{\rm min}$\\
 & & &kpc/$^{\prime\prime}$ 
&mJy&mJy&mJy&arcsec&arcsec\\
\hline
&&&&&&&&\\
3C\,105&S1\footnotemark[1]&0.089&1.642&130$\pm$10&67$\pm$5&45$\pm$5&1.0&0.8\\
 &S2\footnotemark[1]& & &1250$\pm$40&620$\pm$20&460$\pm$15&1.30&1.0\\
 &S3\footnotemark[1]& & &1180$\pm$35&510$\pm$15&320$\pm$12&1.5&0.8\\
 &Ext& & &174$\pm$10&75$\pm$5&50$\pm$3& &  \\
3C\,445&SE\footnotemark[2]&0.0562&1.077&290$\pm$30&98$\pm$15&65$\pm$10&3.5&1.0\\
 &SW\footnotemark[2]& & &220$\pm$25&51$\pm$10&36$\pm$6&1.5&0.5\\
 &Diff\footnotemark[3]& & & & &13.0$\pm$1.1& &  \\
&&&&&&&&\\
\hline
\end{tabular}
\end{center} 
\label{tab_flux_rad}
\end{table*}

\subsection{X-ray observations}

The radio source 
3C\,105 was observed by {\it Chandra} on 2007 December 17 (Obs ID 9299)
during ``The {\it Chandra} 3C Snapshot Survey for Sources with z$<$0.3''
\citep{massaro10}.  
An $\sim$8 ksec 
exposure was obtained with the ACIS-S camera, operating in 
VERY FAINT mode.
The data analysis was performed following the standard
procedures described in the {\it Chandra} Interactive Analysis of
Observations (CIAO) threads and using the CIAO software package v4.2 
(see Massaro et al. 2009 for more details). The {\it Chandra} Calibration
Database (CALDB) version 4.2.2 was used to process all files.  
Level 2 event files were generated using the $acis\_process\_events$ task,
after removing the hot pixels with $acis\_run\_hotpix$.  Events were
filtered for grades 0,2,3,4,6, and we removed pixel randomization.\\
3C\,445 South was observed by {\it Chandra} on 2007 October 18
\citep{perlman10}, ACIS chip S3, with an exposure time of 45.6 ksec. 
The data were retrieved from the archive and
analysed following the same procedure as for 3C\,105 South. This
re-analysis was necessary in order to achieve a proper alignment with
the radio data. \\
We created 3 different flux maps in the soft, medium, and hard X-ray bands
(0.5 -- 1, 1 -- 2, and 2 -- 7 keV, respectively) by dividing the data with
monochromatic exposure maps with nominal energies = 0.8 keV (soft),
1.4 keV (medium), and 4 keV (hard).
Both the exposure maps and the flux maps were regridded to a 
pixel size of 0.25 the size of a native ACIS pixel
(native=0.492$^{\prime\prime}\times0.492^{\prime\prime}$).  To obtain
maps with brightness units of ergs~cm$^{-2}$~s$^{-1}$~pixel$^{-1}$, we
multiplied each event by the nominal energy of its respective band.\\
For 3C\,445 South, we measured a flux density consistent with what reported by
\citet{perlman10}. The flux density was extracted from {\it Chandra}
ACIS-S images in which the hotspot was placed on axis. 
Both hotspots have been detected also by {\it Swift} in the energy range
0.3-10 keV (See Appendix A). This is remarkable given {\it
  Swift}'s survey operation mode and its poor spatial resolution. The
detection level is about 7$\sigma$ and 12$\sigma$ for 3C\,105 South
and 3C\,445 South, respectively. However, given the large {\it Swift}
errors in the counts-to-flux conversion and its low angular
resolution, 
we do not provide any further
flux estimate.\\

\subsection{Image registration}

The alignment between radio and optical images was done by the
superposition of the host galaxies with the nuclear component of the
radio source using the AIPS task LGEOM. This results in a shift of
3.5$^{\prime\prime}$. 
To this purpose, the optical images
were previously brought on the same grid, orientation and coordinate system 
as the radio images by
means of the AIPS task CONV and REGR \citep[see
  also][]{mack09}. 
The final overlay of radio and optical images is
accurate to 0.1$^{\prime\prime}$.\\
For 3C\,105 South the X-ray image has been aligned with the radio one
by comparing the core position. Then, the final
overlay of X-ray contours on the VLT image 
is accurate to 0.1$^{\prime\prime}$. In the case of 3C\,445 the
shape of the nucleus of the galaxy is badly distorted in the {\it Chandra} image
because of its location far off axis of {\it Chandra}.
The alignment was then
performed using three background sources visible both in X-ray and B
band, and located around the hotspot. The achieved accuracy with 
this registration is better than 0.15 arcsec, allowing us to confirm 
a shift of about 2$^{\prime\prime}$ in declination between the X-rays and
B-band emission centroids, the X-ray one being the closest to the core
(Fig. \ref{fig_3c445}).\\

\begin{table*}
\caption{Near infrared, optical flux density and X-ray (0.5 - 7 keV) 
flux of hotspot components. In the case of 3C\,445 the X-ray flux 
is not associated to any of the two
main components. The X-ray flux reported refers the total emission measured
on the whole hotspot region. \newline {\it Note 1}: units in
10$^{-15}$ erg cm$^{-2}$ s$^{-1}$; \newline
{\it Note 2}: the X-ray value, in $\mu$Jy, is from
\citet{perlman10}; \newline
{\it Note 3}: The diffuse emission is inclusive of the SC
component and it is estimated by subtracting from
the total flux density those arising from SW and SE (see Section 5.3).} 
\begin{center}
\begin{tabular}{|c|c|c|c|c|c|c|c|c|c|c|c|}
\hline
Source&Comp.&S$_{\rm K}$&S$_{\rm
  H}$&S$_{\rm J}$&S$_{\rm I}$&S$_{\rm R}$&S$_{\rm B}$&S$_{\rm
  U}$&S$_{\rm I}^{\rm HST}$&S$_{\rm B}^{\rm HST}$&S$_{X}$\\
 & &$\mu$Jy&$\mu$Jy&$\mu$Jy&$\mu$Jy&$\mu$Jy&$\mu$Jy&$\mu$Jy&$\mu$Jy&$\mu$Jy&\\
\hline
&&&&&&&&&&&\\
3C\,105&S1&4.6$\pm$0.9&4.4$\pm$1.1&$<$2.5&- &0.5$\pm$0.1
&0.2$\pm$0.1&-&-&-& 7.5$\pm$2.4\footnotemark[1]\\
 &S2&18.4$\pm$1.4&12.3$\pm$1.1&3.4$\pm$1.0&-&0.7$\pm$0.1&0.2$\pm$0.1&-&-&-&
 $<$2.0\footnotemark[1]\\
 &S3&31.9$\pm$2.8&25.7$\pm$2.9&4.4$\pm$1.8&-&0.9$\pm$0.1&0.3$\pm$0.1&-&-&-&
 3.2$\pm$1.6\footnotemark[1]\\
 &Ext&15.4$\pm$2.0&5.4$\pm$2.0&-& -&0.4$\pm$0.1&0.2$\pm$0.1&-&-&-& -\\
3C\,445&SE&8.0$\pm$1.0&5.6$\pm$2.0&6.0$\pm$1.5&2.0$\pm$0.2&1.3$\pm$0.2&0.7$\pm$0.1&0.5$\pm$0.3&1.7$\pm$0.2&1.5$\pm$0.3&9.38$\times$10$^{-4}$
\footnotemark[2]\\
 &SW&4.6$\pm$1.4&3.6$\pm$1.5&3.0$\pm$0.4&1.7$\pm$0.3&1.4$\pm$0.1&0.7$\pm$0.1&
0.5$\pm$0.2&1.4$\pm$0.1&0.3$\pm$0.1&-\\
 &SC&- &- &- & - &0.8$\pm$0.1&0.6$\pm$0.1&0.4$\pm$0.1&- & - & -\\
 &Diff\footnotemark[3]& -
&2.1$\pm$0.6&3.2$\pm$1.3&1.2$\pm$0.2&1.0$\pm$0.2&0.8$\pm$0.2& - \\
&&&&&&&&&&&\\
\hline
\end{tabular}
\end{center} 
\label{tab_flux_opt}
\end{table*}

\section{Photometry}

To construct the spectral energy distribution (SED) of individual hotspot
components, the flux density at the various wavelengths must be
accurately measured in the same region, avoiding 
contamination from unrelated features. 
To this purpose, we produced a cube where each plane consists of radio
and optical images regridded to the same size and smoothed to the same
resolution. Then the flux
density was derived by means of AIPS task BLSUM which performs an
aperture integration on a selected polygonal region common to all the
images. The values derived in this way were then used to construct the
radio-to-optical SED, and they are reported in
Tables 1 and 2.\\
In addition to the low-resolution approach, 
we derive the hotspot flux densities and
angular sizes on the full resolution images, in order to better
describe the source morphology.

On the radio images, we estimate the flux density of each component 
by means of TVSTAT, which is similar to BLSUM, but instead of working
on an image cube it works on a single image. The angular size was
derived from the lowest contour on the image plane, and it
corresponds to roughly twice the size of the full width half maximum
(FWHM) of a conventional Gaussian covering a similar area. 
In the case of 3C\,105 South, the hotspot components are unresolved
at 1.4 GHz, and we derive the flux density at this frequency by means of
AIPS task JMFIT, which performs a Gaussian fit in the image plane. 
The angular size was measured on the images in
which the components were resolved, i.e. in the case of 3C\,105
South we use the 4.8 and 8.4-GHz images, which provide the same value, while
for 3C\,445 South the components could be reliably resolved in the
image at 8.4 GHz only (Table 1).\\
Full-resolution infrared and optical flux densities of hotspot sub-components 
were measured by means of the IDL-based task ATV using 
a circular aperture centred on each component.
Such values were compared to those derived from the analysis of the
cube and they were found to be within the expected uncertainties.\\
For the X-ray flux we constructed photometric apertures 
to accommodate the {\it Chandra}
point spread function and to include the total extent of the
radio structures.
The background regions, with a total area typically twice that of the
source region, have been selected close to the source, and
centred on a position where other sources or extended structures are
not present. The X-ray flux was measured
in any aperture with only a small correction for the
ratio of the mean energy of the counts within the aperture to the
nominal energy for the band. 
We note that in 3C\,105 South,
the hotspot components 
are well separated (2$^{\prime\prime}$), allowing us to accurately
isolate the corresponding X-ray emission. In
3C\,445 South the X-ray emission is not associated with the two main
components clearly visible in the radio and optical bands, and 
flux was derived by using an aperture large enough to include all of the X-ray
emission extending over the entire hotspot region. Our estimated value
is in agreement with the one reported by \citet{perlman10}.
All X-ray flux densities have been corrected for the
Galactic absorption with the column density N$_H$ =
1.15$\cdot$10$^{21}$cm$^{-2}$ given by
\citet{kalberla05}. 
X-ray fluxes are reported in Table \ref{tab_flux_opt}.\\  

\begin{figure*}
\begin{center}
\includegraphics{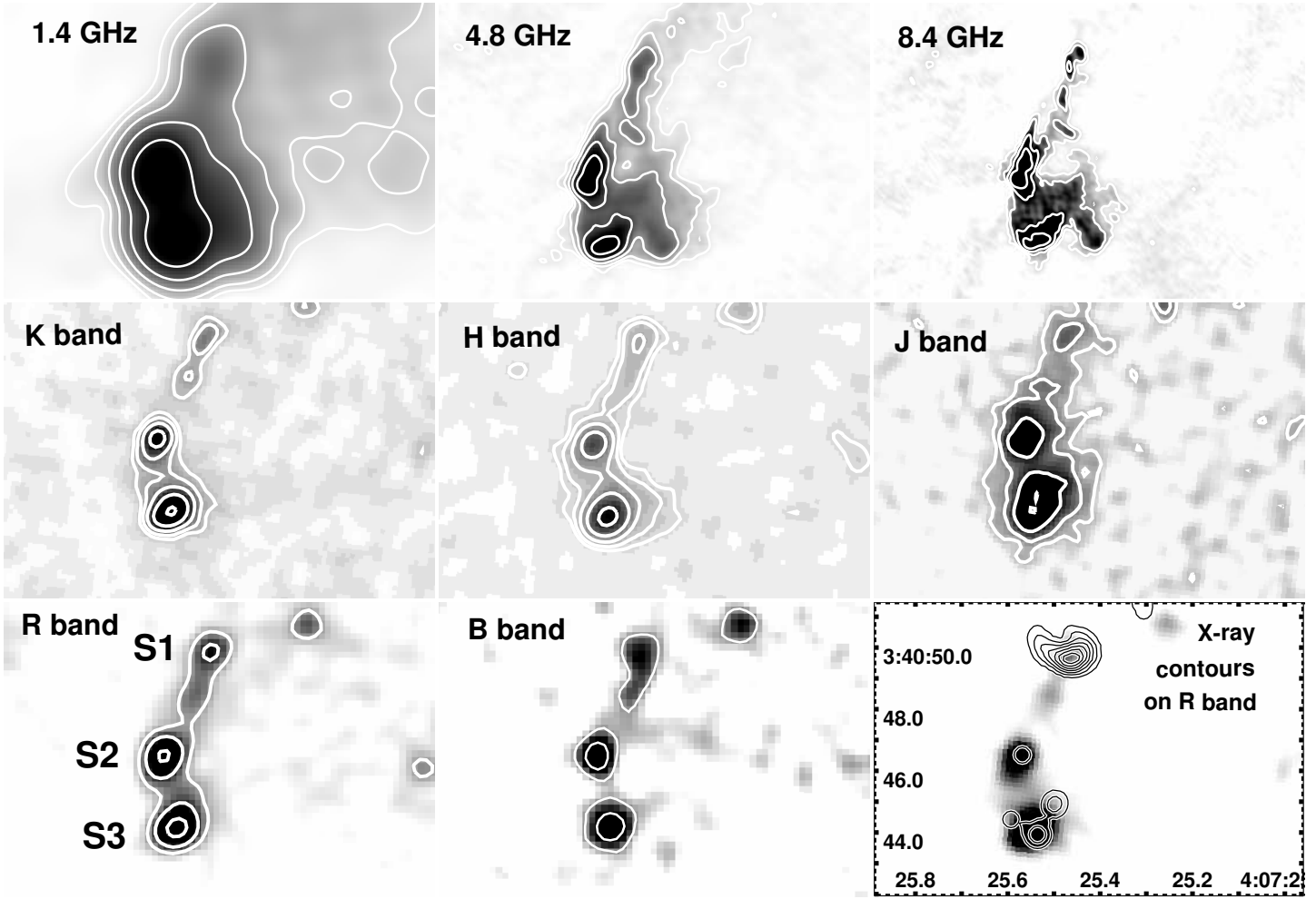}
\vspace{16cm}
\caption{Multifrequency images of 3C\,105 South. From the left to
  right and top to bottom: Radio images at 1.4, 4.8, 8.4 GHz (VLA
  A-array), NIR/optical images in K, H, J, R, B bands (VLT), and X-ray 0.5-7
  keV ({\it Chandra}) contours. Each panel covers 9.5$^{\prime\prime}$
  (15.6 kpc) in DEC and 14$^{\prime\prime}$ (23 kpc) in RA. In the
  radio images the lowest contours are 0.9 mJy/beam at 1.4 GHz, 0.20
  mJy/beam at 4.8 GHz, and 0.18 mJy/beam at 8.4 GHz, and they
  correspond to 3 times the off-source rms noise level measured on the
image plane. Contours increase by a factor of 4. The restoring beam is
1.3$^{\prime\prime}$$\times$1.1$^{\prime\prime}$ at 1.4 GHz,
0.38$^{\prime\prime}$$\times$0.36$^{\prime\prime}$ at 4.8 GHz, and  
0.32$^{\prime\prime}$$\times$0.22$^{\prime\prime}$ at 8.4 GHz. In the
optical images the contour levels are in arbitrary units and increase
by a factor of 2. The FWHM is about 0.4$^{\prime\prime}$,
0.5$^{\prime\prime}$, 0.7$^{\prime\prime}$, 0.6$^{\prime\prime}$,
0.7$^{\prime\prime}$ in K, H, J, R, and B band respectively.
The X-ray contours were generated from an 0.5-7 keV image, smoothed
with a Gaussian of FWHM=0.72$^{\prime\prime}$. Contour levels increase
linearly: 0.02, 0.04, 0.06,.. 0.14 counts per 0.123$^{\prime\prime}$ pixel. 
The X-ray contours
are superposed to the R band image, previously shifted as so to
align with X-ray.} 
\label{fig_3c105}
\end{center}
\end{figure*}

\begin{figure*}
\begin{center}
\includegraphics{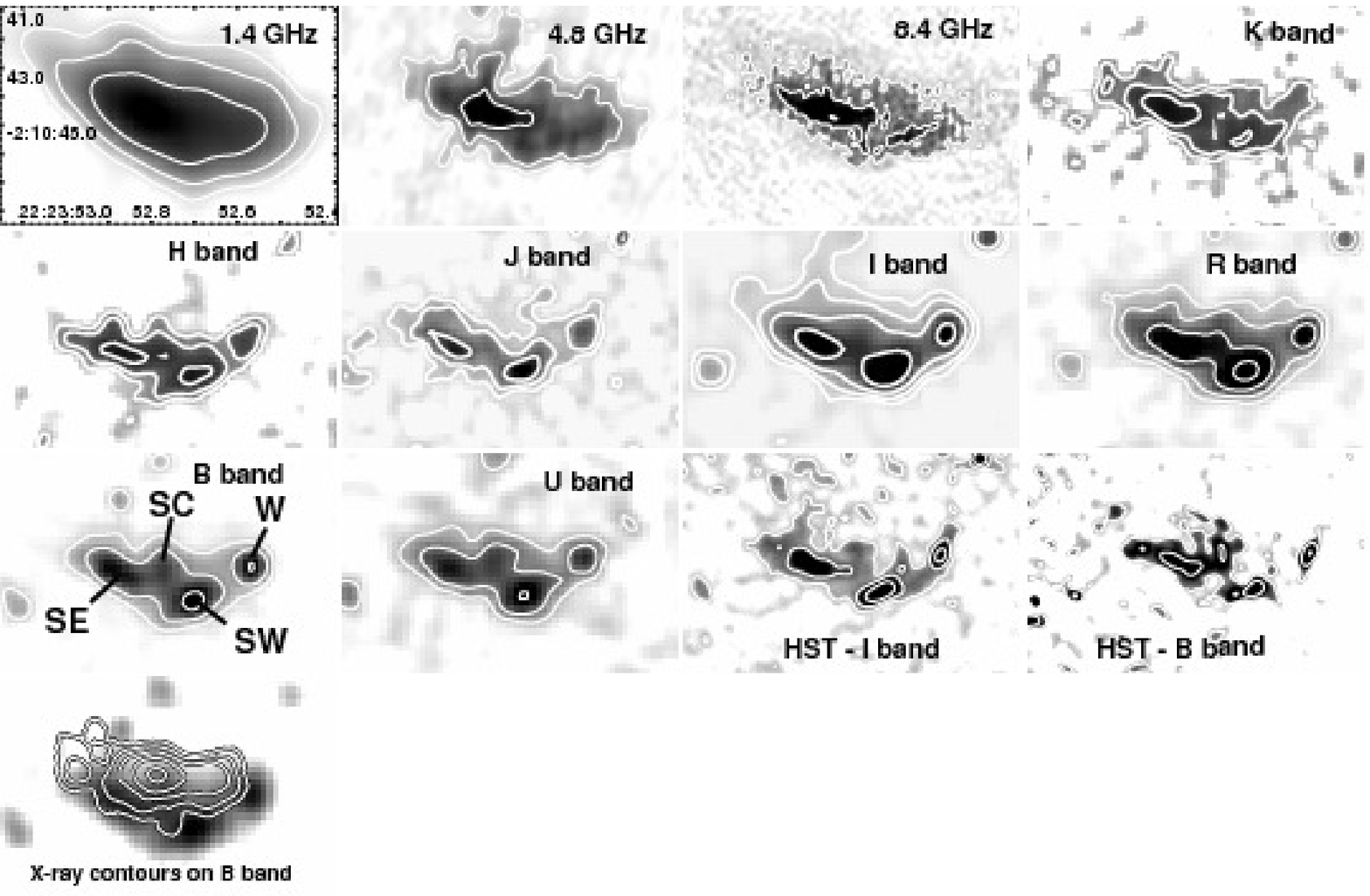}
\vspace{15cm}
\caption{Multifrequency images of 3C\,445 South. From the left to
  right and top to bottom: Radio images at 1.4, 4.8, 8.4 GHz (VLA
  A-array), NIR/optical images in K, H, J, I, R, B, U bands (VLT),
  optical images in I and U bands (HST), 
and X-ray 0.5-7
  keV ({\it Chandra}) contours. Each panel covers 7.3$^{\prime\prime}$
  (7.8 kpc) in DEC and 11.4$^{\prime\prime}$ (12.2 kpc) in RA. In the
  radio images the lowest contours are 1.3 mJy/beam at 1.4 GHz, 0.20
  mJy/beam at 4.8 GHz, and 0.10 mJy/beam at 8.4 GHz, and they
  correspond to 3 times the off-source rms noise level measured on the
image plane. Contours increase by a factor of 4. The restoring beam is
1.43$^{\prime\prime}$$\times$0.96$^{\prime\prime}$ at 1.4 GHz,
0.45$^{\prime\prime}$$\times$0.37$^{\prime\prime}$ at 4.8 GHz, and  
0.24$^{\prime\prime}$$\times$0.21$^{\prime\prime}$ at 8.4 GHz. In the
optical images the contour levels are in arbitrary units and increase
by a factor of 2. The VLT FWHM are 0.7$^{\prime\prime}$,
0.6$^{\prime\prime}$, 0.5$^{\prime\prime}$, 0.7$^{\prime\prime}$,
0.6$^{\prime\prime}$, 0.6$^{\prime\prime}$, 0.7$^{\prime\prime}$,
in K, H, J, I, R, B, and U band respectively. In HST images
each pixel is 0.025$^{\prime\prime}$.
The X-ray contours in the last panel are superposed on the B band
image. They come from an 0.5-7 keV image, smoothed with a Gaussian of
FWHM=0.87$^{\prime\prime}$. Contour levels increase by a factor of 2;
the lowest contour is at a brightness of 0.01 counts per
0.0615$^{\prime\prime}$ pixel.} 
\label{fig_3c445}
\end{center}
\end{figure*}

\section{Morphology}

\subsection{3C\,105 South}

The southern hotspot complex of 3C\,105 shows a curved
structure of about 8$^{\prime\prime}$$\times$4.5$^{\prime\prime}$ 
($\sim$13$\times$7 kpc) in
size. It is dominated by three bright components, all resolved at
radio frequencies,
connected by a low surface brightness emission also visible in
optical and infrared (Fig. \ref{fig_3c105}). 
The central component, labeled S2 in Fig. \ref{fig_3c105}, is the
brightest in radio and, when imaged with high spatial resolution, it
is resolved in two different structures separated by about 1.2 kpc. 
\citet{leahy97} interpreted 
this as the true jet termination hotspot, while S1, with an elongated
structure of (1.6$\times$1.3) kpc and located 5.7 kpc
to the north of S2 is considered as jet emission. The southernmost
component S3,
located about 4.1 kpc from S2, has a resolved structure of
(2.4$\times$1.3) kpc in size, and it is elongated in a direction
perpendicular to the line leading to S2. Its morphology suggests that
S3 is a secondary hotspot similar to 3C\,20 East \citep{cox91}.\\ 
At 1.4 GHz, 
an extended tail
accounting for $S_{\rm 1.4} =$ 608 mJy 
and embedding the jet
is present to the west of the hotspot complex, in agreement with the
structure previously found by \citet{neff95}. 
At higher frequencies the lack of the short spacings prevents
  the detection of such an extended structure, and only a hint of the
jet,
accounting for $S_{\rm 4.8} \sim 70$ mJy, is still
visible at 4.8 GHz. \\
In the optical and NIR the hotspot complex is characterized by
the three main components detected in radio. In NIR and optical, 
the southernmost component S3
is the brightest one, with a radio-to-optical spectral index
  $\alpha_{r-o}=$0.95$\pm$0.10. It displays an elongated
structure rather similar in shape and size to that 
found in radio. It is resolved in all bands with the only exception of 
B band, likely due to the lower spatial resolution 
achieved. Component S1 is resolved in all NIR/optical
bands, showing a tail extending towards S2. Its radio-to-optical
spectral index is $\alpha_{r-o}=$0.95$\pm$0.10.
On the other hand, S2 appears unresolved in all bands, with the
exception of K and H bands, i.e. those with the highest resolution
achieved. In these NIR bands S2 is extended in the southern direction, 
resembling what is observed in radio. Its radio-to-optical
spectral index is $\alpha_{r-o}=$1.05$\pm$0.10.\\
Diffuse emission connecting the main hotspot
components and extending to the southwestern part of the hotspot
complex is
detected in most of the NIR and optical images.\\
In the X-ray band, 
S1 is the brightest component, whereas the emission from 
S3 is very weak (formally detected at only 2$\sigma$ level).
For this reason in the following we will use the
nominal X-ray flux of S3 as a conservative upper limit. 
For component S2 only an upper limit could be set.

\subsection{3C\,445 South}

\begin{figure}
\begin{center}
\includegraphics{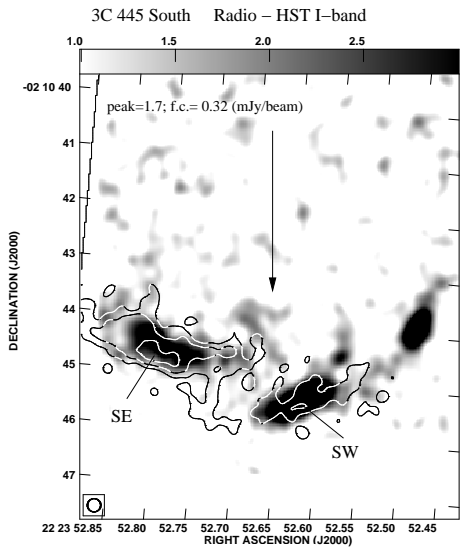}
\vspace{8cm}
\caption{3C\,445 South. 8.4-GHz VLA contours are superimposed on the I-band
HST image.} 
\label{hst_vla}
\end{center}
\end{figure}

The hotspot 3C\,445 South displays an extended east-west structure of
about 9.3$^{\prime\prime}$ $\times$ 2.8$^{\prime\prime}$
(10$\times$3 kpc) in size in radio
(Fig. \ref{fig_3c445}). At 8.4 GHz, the hotspot complex is almost
completely resolved out
and the two main components, clearly visible in
NIR/optical images, are hardly distinguishable.
When imaged with enough resolution, these components display an
arc-shaped structure both in radio and NIR/optical bands,
with sizes of about (3.4$\times$1.5) kpc and
(2.1$\times$1.1) kpc for SE and SW respectively.  
Component SE is elongated in a direction almost perpendicular to the
line leading to the source core, while SW forms an angle of about
-20$^{\circ}$ with the same line.\\ 
In radio and NIR, the 
SE component is the brightest one, with a flux density ratio
SE/SW $\sim$ 1.6, while in the optical both
components have similar flux densities. Both components have a
radio-to-optical spectral index $\alpha_{r-o}=$  0.9$\pm$0.10.
In the optical R-, B-, and U-band 
images a third component (labelled SC in Fig. \ref{fig_3c445}) 
aligned with the jet direction becomes visible between SE
and SW. 
Despite the good resolution and sensitivity of the radio and 
NIR images, SC is not present at such wavelengths.
When imaged with the high
resolution provided by HST, both SE and SW are clearly resolved, and
no compact regions can be identified in the hotspot complex. Trace of
the SC component is seen in the B-band, in agreement
with the VLT images.\\
In the VLA and VLT images, 
the two main components are enshrouded by a diffuse emission, visible
in radio and NIR/optical bands. 
The flux densities of the SE and SW components measured on the HST images
are consistent (within the errors) with those derived on the VLT images.\\ 
The optical component W located about 2.8$^{\prime\prime}$ (3 kpc) 
on the northwestern part of
SW does not have a radio counterpart, as it is clearly shown by
  the superposition of I-band HST and 8.4-GHz VLA images
  (Fig. \ref{hst_vla}), and thus it is considered an
unrelated object, like a background galaxy. Another possibility
  is that this is a synchrotron emitting region where the impact of the
  jet produces very efficient particle acceleration. However, its steep
  optical spectrum ($\alpha \sim 2$ between I and U
  bands, see Section 5.3, Fig. \ref{slope_3c445}) together with the
  absence of detected radio emission disfavour this
  possibility. Future spectroscopic information would further unveil 
  the nature of this optical region.\\
{\it Chandra} observations of 3C\,445 South detected X-ray emission from a
region that extends over 6$^{\prime\prime}$ in the east-west direction
(Fig. \ref{fig_3c445}), and 
it peaks almost in the middle of the hotspot structure, suggesting a
spatial displacement 
between X-ray and radio/NIR/optical emission \citep{perlman10}.\\

\begin{figure}
\begin{center}
\includegraphics{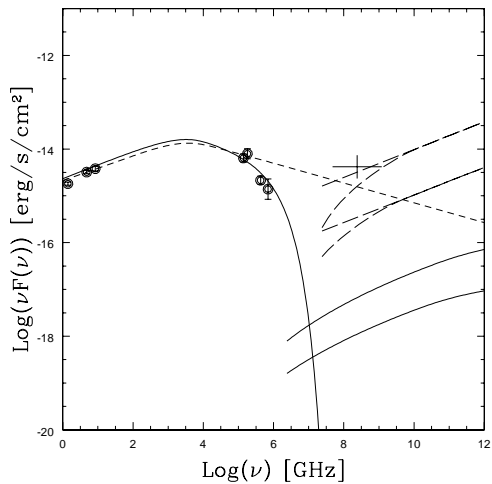}
\vspace{6.5cm}
\caption{The broad-band SED of the northern component, S1, of 
3C\,105 South. The solid lines represent the synchrotron model where
$\nu_{\rm b} =5 \times 10^{12}$ Hz and $\nu_{\rm c} = 2 \times
10^{15}$ Hz, and the SSC
models computed assuming a magnetic field of 50 and 150 $\mu$G. The
short-dashed line represent a synchrotron model where $\nu_{\rm b} = 5
\times 10^{12}$ Hz, and $\nu_{\rm c} = \infty$.
The long-dashed lines represent the IC-CMB models
computed assuming B=16 (and B=32) $\mu$G,
$\Gamma$=6 ($\Gamma=4$), $\theta$=0.1 ($\theta=0.2$) rad,
with or without flattening in the observed synchrotron spectrum
at $\nu<$  60 MHz. The magnetic field is in the rest frame.}
\label{fig_spectra_105n}
\end{center}
\end{figure}

\begin{figure}
\begin{center}
\includegraphics{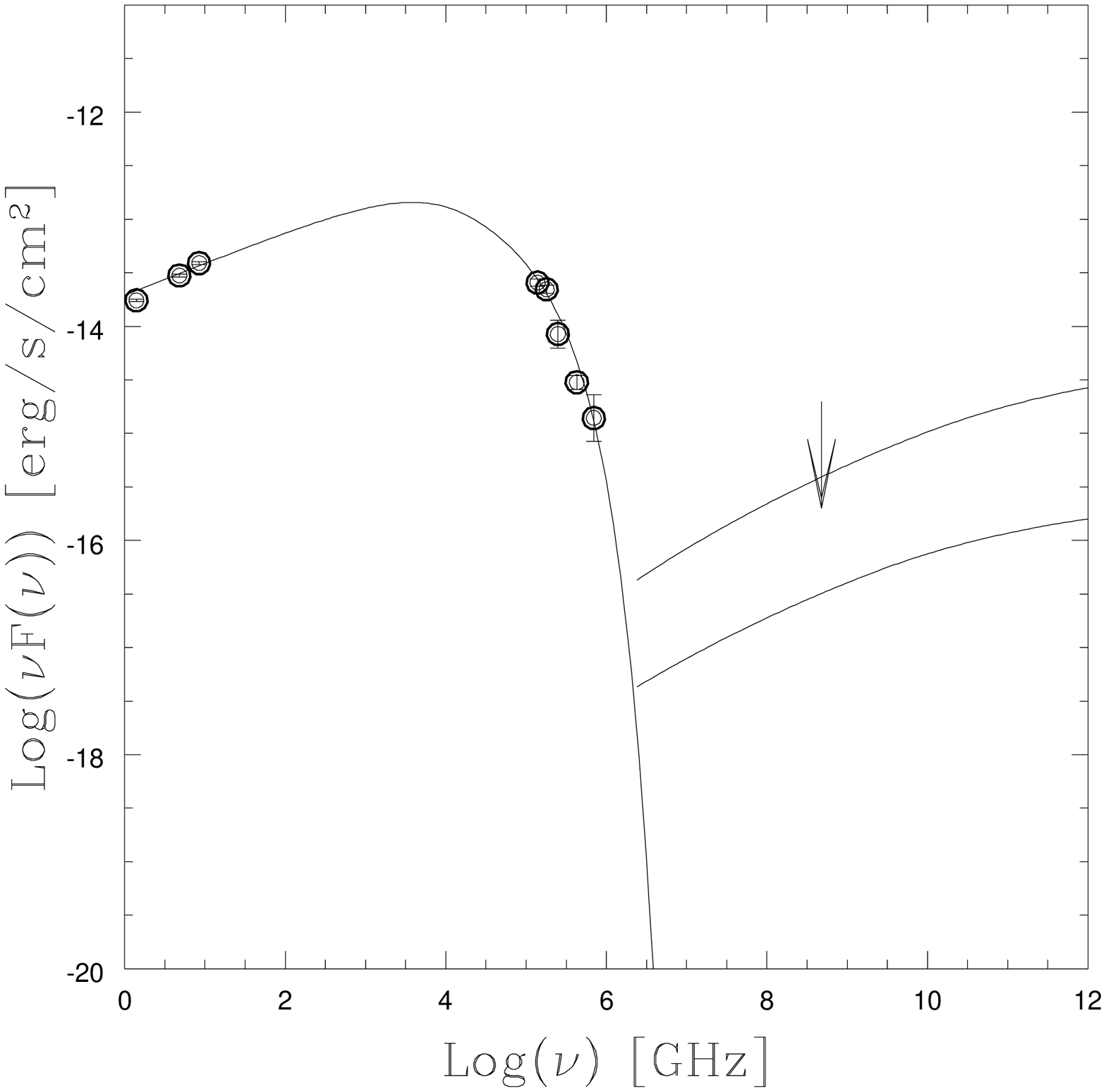}
\vspace{6.5cm}
\caption{The broad-band SED of the central component, S2, of 
3C\,105 South. The solid line represents the synchrotron model where
$\nu_{\rm b} =7.5 \times 10^{12}$ Hz and $\nu_{\rm c} = 3 \times
10^{14}$ Hz, and the SSC
models computed assuming a magnetic field of 50 and 225 $\mu$G. The
arrow indicates the X-ray upper limit.}
\label{fig_spectra_105c}
\end{center}
\end{figure}

\begin{figure}
\begin{center}
\includegraphics{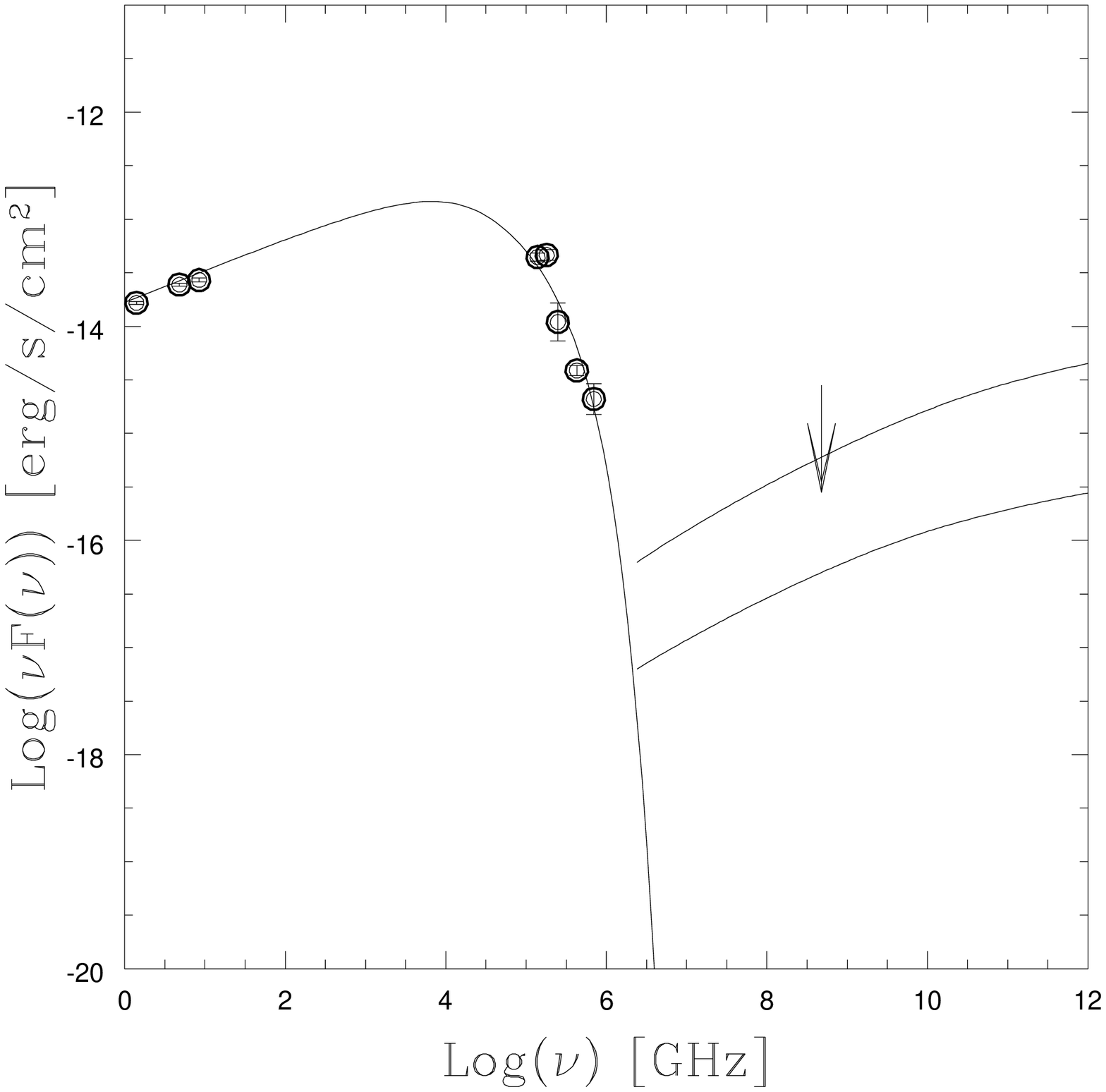}
\vspace{6.5cm}
\caption{The broad-band SED of the southern component, S3, of 
3C\,105 South. The solid line represents the synchrotron model where
$\nu_{\rm b} =1.5 \times 10^{13}$ Hz and $\nu_{\rm c} = 3 \times
10^{14}$ Hz, and the SSC
models computed assuming a magnetic field of 50 and 150 $\mu$G. The
arrow indicates the X-ray upper limit.}
\label{fig_spectra_105s}
\end{center}
\end{figure}

\begin{figure}
\begin{center}
\includegraphics{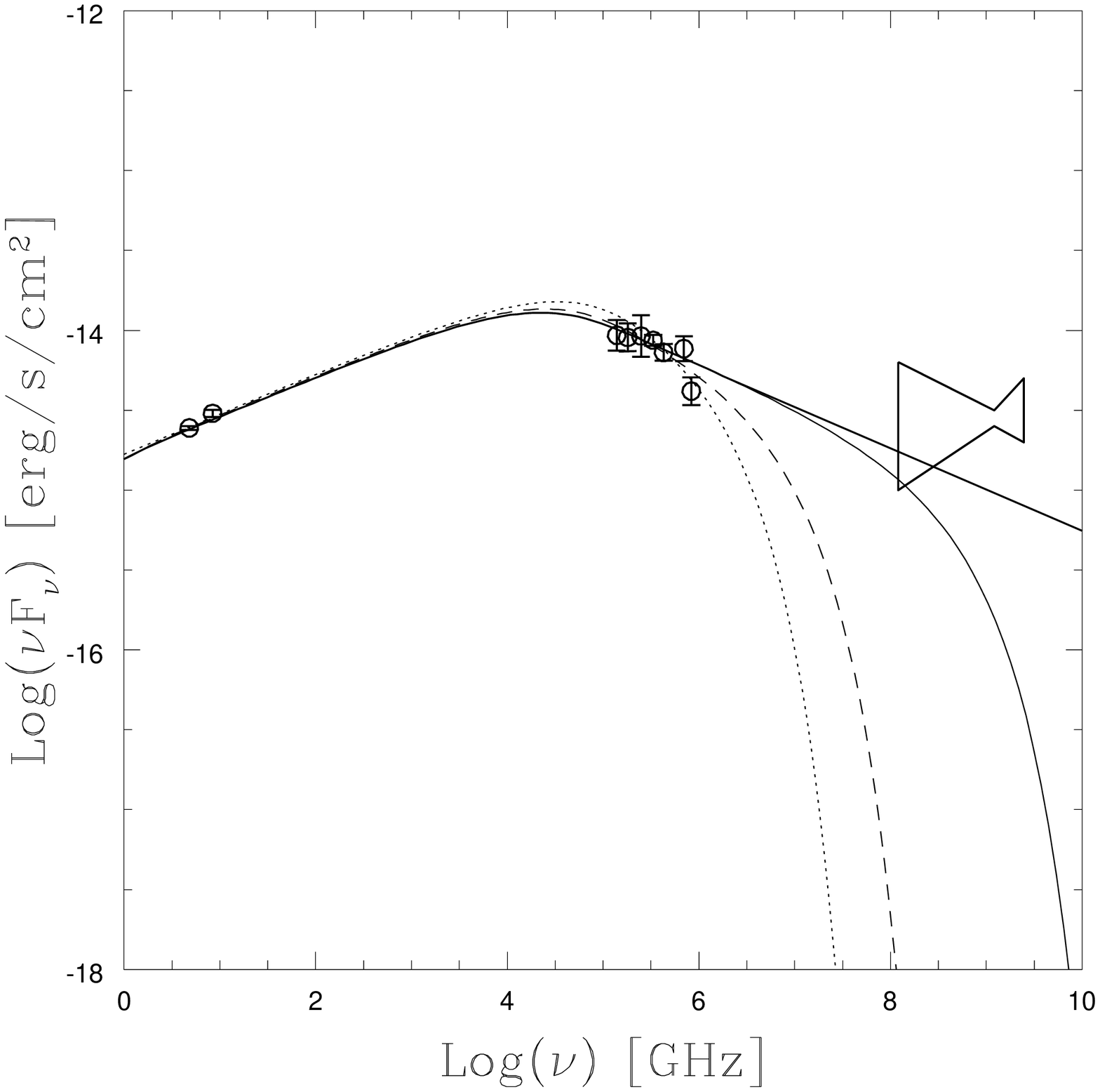}
\vspace{7cm}
\caption{The broad-band SED of the western component, SW, of 
3C\,445 South. The morphology from {\it Chandra} image shows that
X-rays are not associated with the western component.
The synchrotron models assume 
$\nu_{\rm b} =9.4 \times 10^{13}$ Hz and $\nu_{\rm c} = 4.7 \times
10^{15}$ Hz ({\it dotted line}), $\nu_{\rm b} =5.5 \times 10^{13}$ Hz 
and $\nu_{\rm c} = 2.2 \times10^{16}$ Hz ({\it dashed line}), 
$\nu_{\rm b} =4.4 \times 10^{13}$ Hz and $\nu_{\rm c} = 1.8 \times
10^{18}$ Hz ({\it solid line}), $\nu_{\rm b} =4.4 \times 10^{13}$ Hz
and $\nu_{\rm c} = \infty$ ({\it thick solid line}). }
\label{fig_spectra_445w}
\end{center}
\end{figure}

\begin{figure}
\begin{center}
\includegraphics{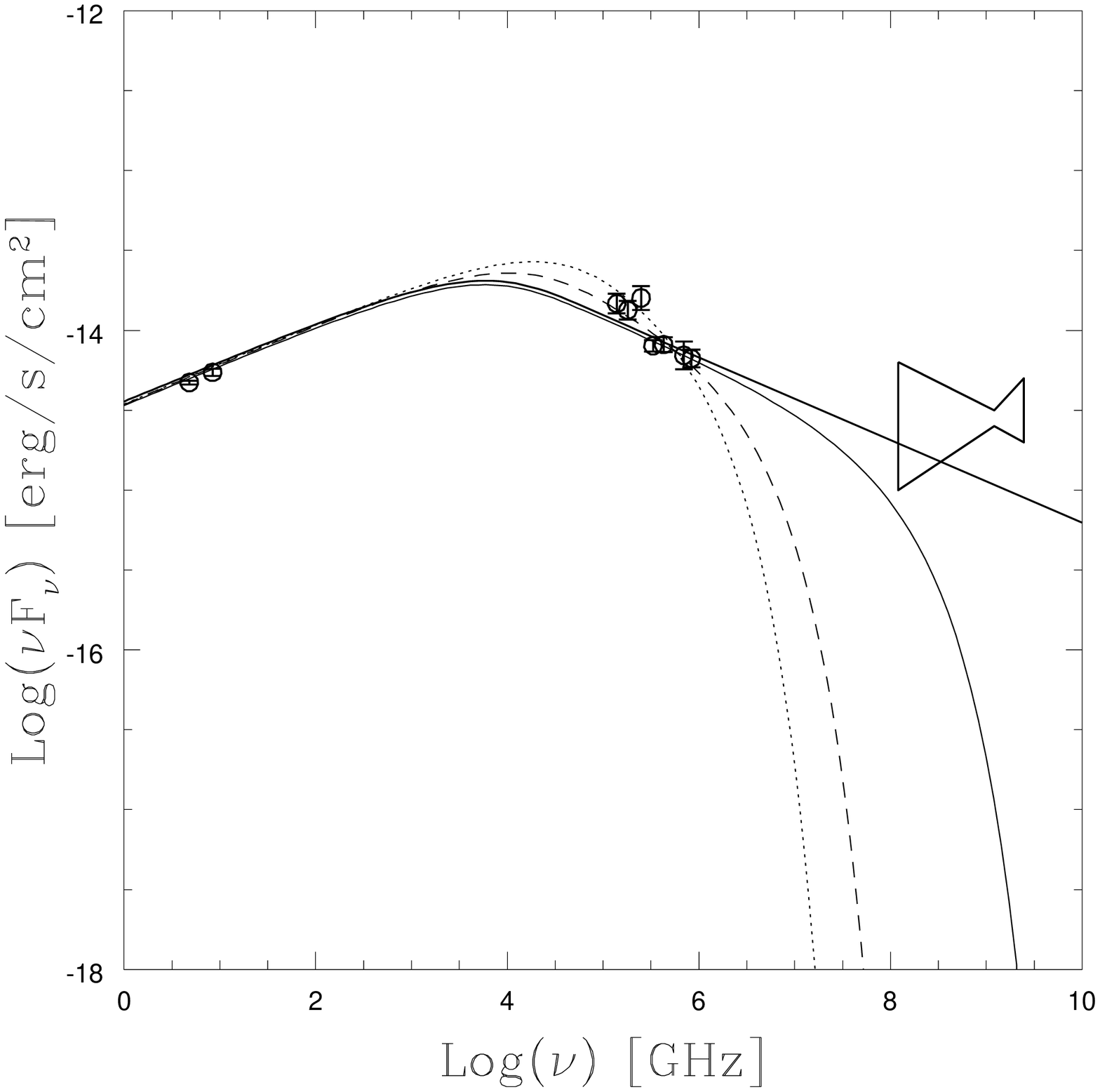}
\vspace{7cm}
\caption{The broad-band SED of the eastern component, SE, of 
3C\,445 South. The morphology from {\it Chandra} image shows that
X-rays are not associated with the eastern component.
The synchrotron models assume 
$\nu_{\rm b} =5.2 \times 10^{13}$ Hz and $\nu_{\rm c} = 2.6 \times
10^{15}$ Hz ({\it dotted line}), $\nu_{\rm b} =2.4 \times 10^{13}$ Hz 
and $\nu_{\rm c} = 9.4 \times10^{15}$ Hz ({\it dashed line}), 
$\nu_{\rm b} =1.2 \times 10^{13}$ Hz and $\nu_{\rm c} = 4.7 \times
10^{17}$ Hz ({\it solid line}), $\nu_{\rm b} =1.2 \times 10^{13}$ Hz
and $\nu_{\rm c} = \infty$ ({\it thick solid line}). }
\label{fig_spectra_445e}
\end{center}
\end{figure}

\begin{figure}
\begin{center}
\includegraphics{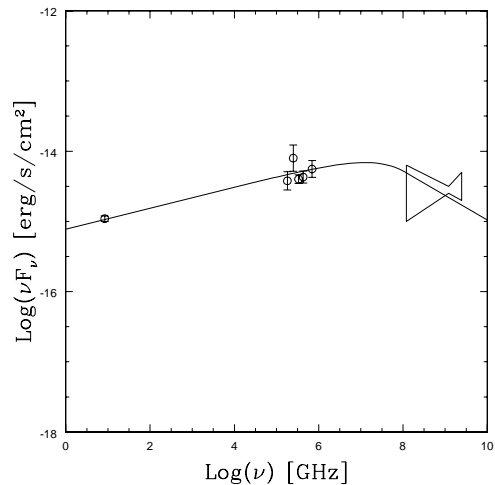}
\vspace{7cm}
\caption{The broad-band SED of the diffuse emission (see text) of 
3C\,445 South. The morphology from the {\it Chandra} image does not
allow us to firmly exclude a connection between the X-rays and the
diffuse (including SC component) emission.
The synchrotron model assumes
$\nu_{\rm b} =8 \times 10^{16}$ Hz, $\nu_{\rm c} \gg \nu_{\rm b}$ and $p$=2.7.}
\label{fig_spectra_445diff}
\end{center}
\end{figure}

\section{Spectral energy distribution}

\begin{table}
\caption{Synchrotron parameters. Column 1: Hotspot; Column 2:
  component; Columns 3, 4: spectral index and break frequency as 
derived from the fit to the radio-to-optical SED (Section 5.1); 
Column 5: equipartition magnetic field, computed following the
approach presented in Brunetti et al. (2002); Column 6: radiative age
computed using Eq. 2.} 
\begin{center}
\begin{tabular}{|c|c|c|c|c|c|}
\hline
Source&Comp.&$\alpha$&$\nu_{\rm b}$&$B_{\rm eq}$&$t_{\rm rad}$\\
 & & &10$^{13}$ Hz&$\mu$G&yr\\
\hline
&&&&&\\
3C\,105&S1&0.8&0.50&150&12\\
       &S2&0.8&0.75&290&4\\
       &S3&0.8&1.5&270&3\\
3C\,445&SE&0.75&5.2&60&15\\
       &SW&0.75&9.4&50&15\\
&&&&&\\
\hline
\end{tabular}
\label{tab_fit}
\end{center}
\end{table}

\subsection{The broad-band energy distribution}

We model the broad band energy distribution, from radio to optical, of the
hotspot regions in order to determine the mechanisms at the basis of
the emission. The comparison between the model expectation in the
X-rays and {\it Chandra} data sets additional
constraints.
In the adopted models, the
hotspot components are described by homogeneous spheres with constant magnetic
field and constant properties of the relativistic electron
populations. 
The spectral energy distributions of the emitting electrons are modelled
assuming the formalism described in \citet{gb02}. According to
  this model a population of seed electrons (with $\gamma \leq
  \gamma_{*}$) is accelerated at the shock and is injected in the
  downstream region with a spectrum dN($\gamma$)/dt $\propto$
  $\gamma^{-p}$, for $\gamma_{*} < \gamma < \gamma_{c}$, 
  $\gamma_{c}$ being the maximum energy of the electrons accelerated at the
  shock. Electrons accelerated at the shock are advected in the
  downstream region and age due to radiative losses. Based on
  \citet{gb02}, the volume integrated spectrum of the electron
  population in the downstream region of size $L \sim T v_{\rm adv}$
  ($T$ and $v_{\rm adv}$ being the age and the advection velocity of
  the downstream region) is given by either a steep power-law $N$($\gamma$)
  $\propto \gamma^{-(p+1)}$ for $\gamma_{b} < \gamma < \gamma_{c}$,
  where $\gamma_{b}$ is the maximum energy of the ``oldest'' electrons
in the downstream region, or by $N$($\gamma$)
  $\propto \gamma^{-p}$ for $\gamma_{*} < \gamma < \gamma_{b}$, or by
a flatter shape for $\gamma_{\rm low} < \gamma < \gamma_{*}$, where
$\gamma_{\rm low}$ is the minimum energy of electrons accelerated at
the shock.\\
As the first step we fit the SED in the
radio-NIR-optical regimes with a synchrotron model, and we derive the
relevant parameters of the synchrotron spectrum  
(injection spectrum $\alpha$, break frequency
$\nu_{\rm b}$, cut-off frequency $\nu_{\rm c}$) 
and the slope of the energy distribution of the electron population
as injected at the shock (p $= 2 \alpha +1$). Since hotspots have
  spectra with injection slope $\alpha$ ranging between 0.5 and 1 (as
  a reference, the classical value from the diffuse particle
  acceleration at strong shocks is $\alpha = 0.5$, e.g. Meisenheimer
  et al. 1997), we decided to consider the injection spectral index as
  a free parameter.
Such constraints allow us to determine the spectrum of the emitting
electrons (normalization, break and cut-off
energy), once the magnetic field strength has been assumed, and
to calculate the emission from either synchrotron-self-Compton (SSC) 
or inverse-Compton scattering of the cosmic background radiation 
(IC-CMB) expected from the hotspot (or jet)
region \citep[following][]{gb02}.
Models described in \citet{gb02} take also into account the boosting
effects arising from a hotspot/jet that is moving at relativistic
speeds and oriented at a given angle with respect to our line of
sight.\\

\subsection{3C\,105 South}

In Figures \ref{fig_spectra_105n} to \ref{fig_spectra_105s} 
we show the SED 
from the radio band to high energy emission measured
for the hotspot components of 3C\,105 South, together with the model fits.
Synchrotron models with an injection spectral index $\alpha$=0.8
provide an adequate representation of the SED
of the central and southern components of
3C\,105 South, with break frequencies ranging from
5$\times$10$^{12}$ to 1.5$\times$10$^{13}$ W/Hz, while the cutoff
frequencies are between 3$\times$10$^{14}$ and 2$\times$10$^{15}$ W/Hz.
In both components, the upper limit to the X-ray emission does
not allow us to constrain the validity of the SSC model
(Figs. \ref{fig_spectra_105c} and \ref{fig_spectra_105s}).  
On the other hand, the northern component of 3C\,105 shows a prominent
X-ray emission.
A synchrotron model (dashed line in Fig. \ref{fig_spectra_105n}) may
fit quite reasonably the radio, NIR and X-ray emission, but
it completely fails in reproducing
the optical data. An additional contribution of the SSC 
is not a viable option since it requires a 
magnetic field much smaller than that
obtained assuming equipartition (see Section 5.4) (solid lines), and 
implying an
unreasonably large energy budget.
On the other hand, the high energy emission is well modelled by
IC-CMB \citep[e.g.][]{tavecchio00,celotti01} 
where the CMB photons are scattered by relativistic electrons with
Lorentz factor $\Gamma \sim 6$, and $\theta$=5$^{\circ}$ with a
magnetic field of 16 $\mu$G. 
This model
implies that boosting effects play an important role in the X-ray emission
of this component, suggesting that S1 is more likely a relativistic
knot in the jet, rather than a hotspot feature. The weakness of
  this interpretation is that 3C\,105 is a NLRG and its jets are
  expected to form a large angle with our line of
  sight.
Alternatively,
  the X-ray emission may be synchrotron from a different population of
electrons, as suggested in the case of the jet in 3C\,273 (Jester et
al. 2007).\\ 

\subsection{3C\,445 South}

\begin{figure}
\begin{center}
\includegraphics{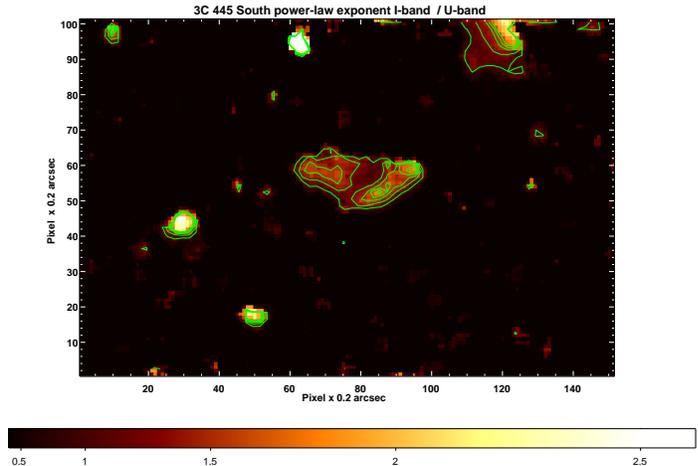}
\vspace{6.5cm}
\caption{Power-law spectral index map for 3C\,445 South determined from FORS
I-band and FORS U-band. Contours are
1, 1.3, 1.5, 1.6, 1.7. First contour is 3 sigma.}
\label{slope_3c445}
\end{center}
\end{figure}

The analysis of the southern hotspot of 3C\,445 
as a single unresolved component was carried out in previous work by
\citet{aprieto02,mack09,perlman10}. In this new analysis, 
the high spatial resolution and
multiwavelength VLT and HST data of 3C\,445 South allow us 
to study the SED of each
component separately in order to investigate in more detail the
mechanisms at work across the hotspot region.
In Figures \ref{fig_spectra_445w} to \ref{fig_spectra_445diff} 
we show the SED 
from the radio band to high energy emission measured
for the components of 3C\,445 South, together with the model fits.
We must note that at 1.4 GHz the resolution is not
sufficient to reliably separate the contribution from the two
main components. For this reason,
we do not consider the flux density at this frequency in constructing
the SED. The X-ray emission
(Fig. \ref{fig_3c445}) is misaligned with respect to the radio-NIR-optical
position. For this reason, on the SED of
both components (Figs. \ref{fig_spectra_445w} and
\ref{fig_spectra_445e}) we plot the total X-ray flux which must be
considered an upper limit. 
For the components of 3C\,445 South the synchrotron models with
$\alpha$=0.75 reasonably
fit the data, providing break frequencies in the range of 10$^{13}$
and $10^{14}$ W/Hz, and cutoff frequencies from 10$^{15}$ Hz and
10$^{18}$ Hz.\\
Both the morphology (Fig. \ref{fig_3c445}) and the SED
(Figs. \ref{fig_spectra_445w} and 
\ref{fig_spectra_445e}) indicate that the bulk of {\it Chandra} X-ray
emission detected in 3C\,445 is not due to synchrotron emission from
the two components (Section 6).\\
As discussed in Section 4.2, diffuse IR and optical emission
  surrounds the two components SE and SW of 3C\,445 South, and a third
component, SC, becomes apparent in the optical. We attempt to evaluate the
spectral properties of the diffuse emission (including component
SC). When possible, depending on statistics, we subtract from the
total flux density of the hotspot, the contribution 
arising from the two main
components, obtaining in this way the SED of the diffuse emission
(inclusive of SC component) of
3C\,445 South. In the image we also plot the total X-ray flux. 
As expected the emission has a hard spectrum ($\alpha
\sim 0.85$) without evidence of a break up to the optical band,
10$^{15}$ Hz $<$ $\nu_{b}$ $\leq$ 8$\times$10$^{16}$ Hz. We also note
that this hard component may represent a significant contribution of
the observed X-ray emission, although the X-ray peak appears shifted
($\sim$ 1$^{\prime\prime}$) from the SC component.
Due to the extended nature of the emission in this hotspot, we
  created a
power-law spectral index map 
illustrating the change of the spectral index $\alpha$
across the hotspot region (Fig. \ref{slope_3c445}). 
The spectral energy distributions presented in
Figs. \ref{fig_spectra_445w}, \ref{fig_spectra_445e}, and
\ref{fig_spectra_445diff} show the
curvature of the integrated spectrum for the main
components and the diffuse emission (see Section 5.1). 
The spectral map in Fig. \ref{slope_3c445} attempts to provide
complementary information on the spectral slope for the diffuse
inter-knot emission. Extracting these maps using the largest 
possible frequency range is complicated as it implies combining images 
from different instruments with different scale sampling, 
noise pattern, etc. These effects sum up to produce very 
low contrast maps given the weakness of the hotspot signal. 
To minimise these effects it was decided to extract the slope maps from
the optical and -IR images only.\\
The spectral index map between I- and U-band (Fig. \ref{slope_3c445}) shows 
two sharp edges, at the SW and SE components, with the highest value 
$\alpha \sim  1.5 $. 
Between these two main regions there 
is diffuse emission that is clearly seen
in the I-/U-band spectral index map. The slope of this 
component is flatter than that of the two main regions 
and rather uniform all over the hotspot, with $\alpha \sim 1$.\\

\subsection{Physical parameters}

We compute the magnetic field of each hotspot component by 
assuming minimum energy conditions,
corresponding to equipartition of energy between radiating
particles and magnetic field, 
and following the approach by \citet{gb97}.
We assume for the hotspot components an ellipsoidal volume $V$ with a
filling factor $\phi$=1 (i.e. the volume is fully and homogeneously
filled by relativistic plasma). 
The volume $V$ is computed by means:\\

\begin{equation}
V = \frac{\pi}{6} d_{\min}^{2} d_{\max}
\end{equation}

\noindent where d$_{\min}$ and d$_{\max}$ are the linear size of the
minor and major axis, respectively.  
We consider $\gamma_{\rm min} =$100, 
and we assume that the energy densities
of protons and electrons are equal. 
We find equipartition 
magnetic fields ranging from $\sim$ 50 - 290 $\mu$G (Table
\ref{tab_fit}) that is 
lower than those 
inferred in high-power radio hotspots 
which range from $\sim$ 250 to 650 $\mu$G
\citep{meise97, cheung05}. 
Remarkably, if we compare these results with those from \citet{mack09},
we see that in 3C\,445 South the value
computed considering the entire source volume is similar to those obtained in
its individual sub-components, suggesting that compact
and well-separated emitting regions are not present in the hotspot volume. 
On the other hand, the magnetic field
averaged over the whole 3C\,105 South hotspot complex is much smaller
than those derived in its sub-components.\\
In the presence of such low magnetic fields 
high-energy electrons may have longer radiative lifetime than
in high-power radio hotspots. 
The radiative age $t_{\rm rad}$ is related to the
magnetic field and the break frequency by\footnote{The magnetic field
  energy density in these hotpots are at least an order of magnitude
  higher than the energy density of the cosmic microwave background
  (CMB) radiation. Inverse Compton
  losses due to scattering of CMB photons
  are negligible.}:\\

\begin{equation}
t_{\rm rad} = 1610 \; B^{-3/2} \nu_{b}^{-1/2} (1+z)^{-1/2}
\label{eq_trad}
\end{equation}

\noindent where B is in $\mu$G, $\nu_{b}$ in GHz and $t_{\rm rad}$ in
10$^{3}$ yr. If in Eq. \ref{eq_trad} we assume the equipartition
magnetic field 
we find that the radiative ages are just a few years (Table 3). 
As the hotspots
extend over kpc distances, it is indicative that a very efficient
re-acceleration mechanism is operating in a similar way over
the entire hotspot region.\\

\section{Discussion}

The detection of diffuse optical emission occurring well outside
the main shock region and distributed over a large fraction of the
whole kpc-scale hotspot structure is somewhat surprising. Deep optical
observations pointed out that this is a rather common phenomenon
detected in about a dozen hotspots \citep[e.g.][]{mack09,cheung05,thomson95}.
First-order Fermi
acceleration alone cannot explain optical emission extending on kpc
scale and additional efficient mechanisms taking place away from the
main shock region should be considered, 
unless projection effects play an important role in smearing compact
regions where acceleration is still occurring.\\
Theoretically, we can consider several scenarios that are able to reproduce
the observed extended structures.
(1) One possibility is that a very wide jet, with a size comparable to the 
hotspot region, impacts simultaneously into various locations across the
hotspot generating a complex shocked region that defines an arc-shaped
structure. This, combined with projection effects may explain a
wide (projected) emitting region.
(2) Another possibility is a narrow jet that impacts into the hotspot in a
small region where electrons are accelerated at a strong shock. In this
case the accelerated particles are then transported upstream 
in the hotspot
volume where they are continuously re-accelerated by stochastic mechanisms,
likely due to turbulence generated by the jet and shock itself.
(3) Finally, extended emission may be explained 
by the ``dentist's drill'' scenario, in which the jet impacts into
the hotspot region in different locations at different times.  \\
The peculiar morphology and the rather high NIR/optical luminosity 
of 3C\,105 South and 3C\,445 South, makes 
these hotspots ideal targets to investigate the
nature of extended diffuse emission. \\
In 3C\,105 South, the detection of optical emission in both 
primary and
secondary hotspots implies that in these regions there is a continuous 
re-acceleration of particles. The secondary hotspot S3 could be interpreted
as a splatter-spot from material accelerated in the primary one, S2
\citep{williams85}.  
Both the alignment and the distance between these
components exclude the jet drilling scenario: 
the light time between the two components is more than 10$^{4}$ years,
i.e. much longer than their radiative time (Table 3), suggesting that
acceleration is taking place in both S2 and S3
simultaneously. The secondary hotspot S3 shows some
elongation, always in the same direction, in all the radio and optical
images with adequate spatial resolution. 
This elongation is expected in a splatter-spot and it
follows the structure of the shock generated by the impact of the 
outflow from the
primary upon the cocoon wall. \\
This scenario, able to explain the presence of optical emission from two
bright and distant components, fails in reproducing the diffuse
optical emission enshrouding the main features, and the extended
tail. In this case, an additional contribution from stochastic mechanisms
caused by turbulence in the downstream region is necessary. 
Although this acceleration mechanism is in general less
efficient than Fermi-I processes, the (radiative) energy losses of
particles are smaller in the 
presence of low magnetic fields, such as those in between S2 and S3, 
(potentially)
allowing stochastic mechanisms to maintain electrons at high energies. \\
In 3C 445 South the observational picture is complex.
The optical images of 3C 445 South show a spectacular 10-kpc arc-shape
structure. High resolution HST images allow a further step since they 
resolve this structure in two elongated components enshrouded by diffuse 
emission. 
These components may mark the regions where a ``dentist's drill'' jet
impacts on the ambient medium, representing the most recent episode
of shock acceleration due to the jet impact. On the other hand, they
could simply trace the locations 
of higher particle-acceleration efficiency from a wide/complex 
interaction between the jet and the ambient medium.
However, the transverse extension, about 1 kpc, of the two elongated 
components is much larger than what is derived if the relativistic
particles, accelerated at the shock, age in the downstream region (provided 
that the hotspot advances at typical speeds of 0.05-0.1$c$).
Furthermore, the diffuse optical emission on larger scale
suggests the presence of additional, complex, acceleration mechanisms,
such as stochastic processes, 
able to keep particle re-acceleration ongoing in the 
hotspot region.
The detection of X-ray emission with {\it Chandra} 
adds a new grade of complexity. This emission and its displacement 
are interpreted by \citet{perlman10} as due to IC-CMB originating in
the fast part of the decelerating flow. Their model requires that the
angle between the jet velocity and the observer's line  
of sight is small. 
However, 3C\,445 is a
  classical double radio galaxy and the jet should form a large angle
  with the line of sight (see also Perlman et al. 2010).
On the other hand, 
we suggest that the X-ray/optical offset might be the outcome of 
  ongoing efficient particle acceleration occurring in the hotspot
  region. An evidence supporting this interpretation
may reside on the faint and diffuse blob
seen in U- and B-bands (labelled SC in Fig. \ref{fig_3c445}) just
about 1$^{\prime\prime}$ downstream the X-ray peak. The surface
brightness of this
component decreases rapidly as the frequency decreases, as it is shown in
Fig. \ref{fig_3c445}: well-detected in U- and
B-bands, marginally visible in I-band, and absent at NIR and radio
wavelengths. The SED of the diffuse hotspot emission (including SC
component and excluding SW and SE) is consistent with synchrotron
emission with a break at high frequencies, 10$^{15}$ Hz $<$ $\nu_{b}$
$\leq$ 8$\times$10$^{16}$ Hz, and may significantly contribute to the
observed X-ray flux. 
Such a hard spectrum is in agreement with (i) a
very recent episode of particle acceleration (the radiative cooling time of the
emitting particles being 10$^{2}$-10$^{3}$ yr); (ii) efficient
spatially-distributed acceleration processes,
similar to the scenario proposed for the western hotspot of Pictor A 
(Tingay et al. 2008, see their Fig.5). \\

\section{Conclusions}

We presented a multi-band, high spatial resolution study of the
hotspot regions in two nearby radio galaxies,
namely 3C\,105 South and 3C\,445 South, on the basis of 
radio VLA, NIR/optical VLT and HST, and X-ray {\it Chandra}
observations. At the sub-arcsec resolution achieved at radio and
optical wavelengths, both hotspots display
multiple resolved components connected by diffuse emission detected
also in optical. The hotspot region in 3C\,105 resolves
in three major components: a primary hotspot, unresolved and aligned
with the jet direction, and a secondary hotspot, elongated in shape,
and interpreted as a splatter-spot arising from continuous outflow of
particles from the primary. 
Such a feature, together with the extremely short
radiative ages of the electron populations emitting in the optical,
indicates that the jet has been impacting
almost in the same position for a long period, making the
drilling jet scenario unrealistic. 
The detection of an excess of X-ray
emission from the northern component of 3C\,105 South 
suggests that this region is likely a relativistic knot in the jet
rather than a genuine hotspot feature.
The optical diffuse emission enshrouding
the main components and extending towards the tail can
be explained possibly assuming additional stochastic mechanisms
taking place across the whole hotspot region.\\ 
In the case of 3C\,445 South the optical observations probe a scenario 
where the interaction between jet and the ambient medium is very complex.
Two optical components pinpointed by HST observations mark either the locations 
where particle acceleration is most efficient or the remnants of the most
recent episodes of acceleration.
Although projection effects may play an important role, the morphology 
and the spatial extension of the diffuse optical emission suggest that
particle accelerations, such as stochastic mechanisms, 
add to the standard shock acceleration
in the hotspot region.
The X-rays detected by {\it Chandra} cannot be the counterpart at higher
energies of the two main components. It might be due to
IC-CMB from the fast part of a decelerating flow.
Alternatively the X-rays could pinpoint synchrotron emission from
recent episodes of efficient particle acceleration occurring in the
whole hotspot region, similarly to what proposed in other hotspots, that
would make the scenario even more complex.
A possible evidence supporting this scenario comes from the
  hard spectrum of the diffuse hotspot emission and from the
  appearance of a new component (SC) in the optical images.

\section*{Acknowledgment}
We thank the anonymous referee for the valuable suggestions that improved the manuscript. 
F.M. acknowledges the Foundation BLANCEFLOR Boncompagni-Ludovisi, n'ee
Bildt for the grant awarded him in 2010 to support his research.
The VLA is operated by the US 
National Radio Astronomy Observatory which is a facility of the National
Science Foundation operated under cooperative agreement by Associated
Universities, Inc. This work has made use of the NASA/IPAC
Extragalactic Database NED which is operated by the JPL, Californian
Institute of Technology, under contract with the National Aeronautics
and Space Administration. This research has made used of SAOImage DS9,
developed by the Smithsonian Astrophysical Observatory (SAO). Part of
this work is based on archival data, software or on-line services
provided by ASI Science Data Center (ASDC). The work at SAO is
supported by supported by NASA-GRANT GO8-9114A. 
We acknowledge the use of public data from
the Swift data archive. This research has made use of software
provided by the Chandra X-ray Center (CXC) in the application packages
CIAO and ChIPS.

\appendix
\section{{\it Swift} images}

Both the radio hotspot 3C\,105 South and 3C\,445 South have been
detected by {\it Swift} in the energy range 0.3-10 keV. \\
The reduction procedure for {\it Swift} data follows 
that described in \citet{massaro08}.
In the following we report only the basic details.\\

\begin{figure}
\begin{center}
\includegraphics{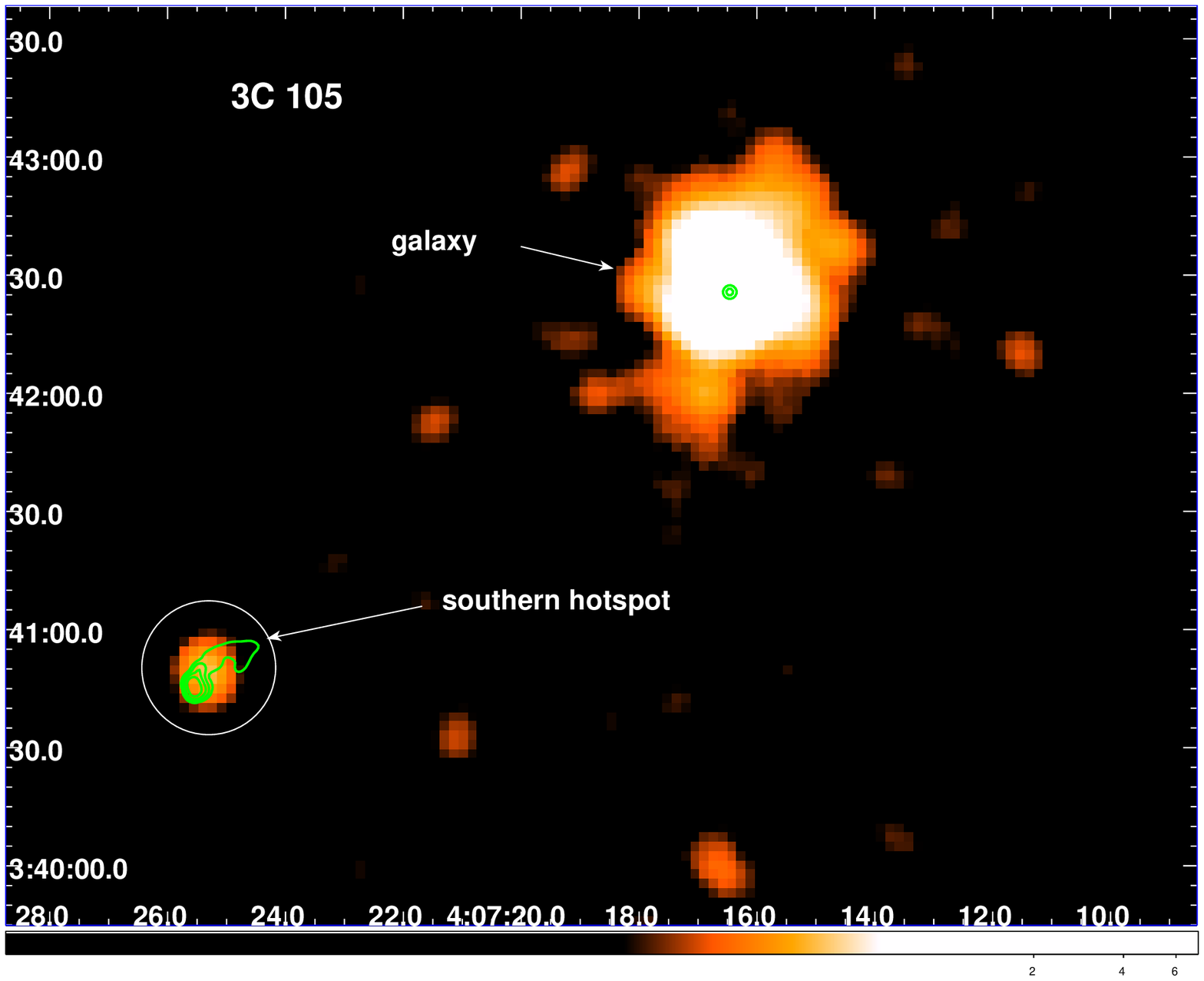}
\includegraphics{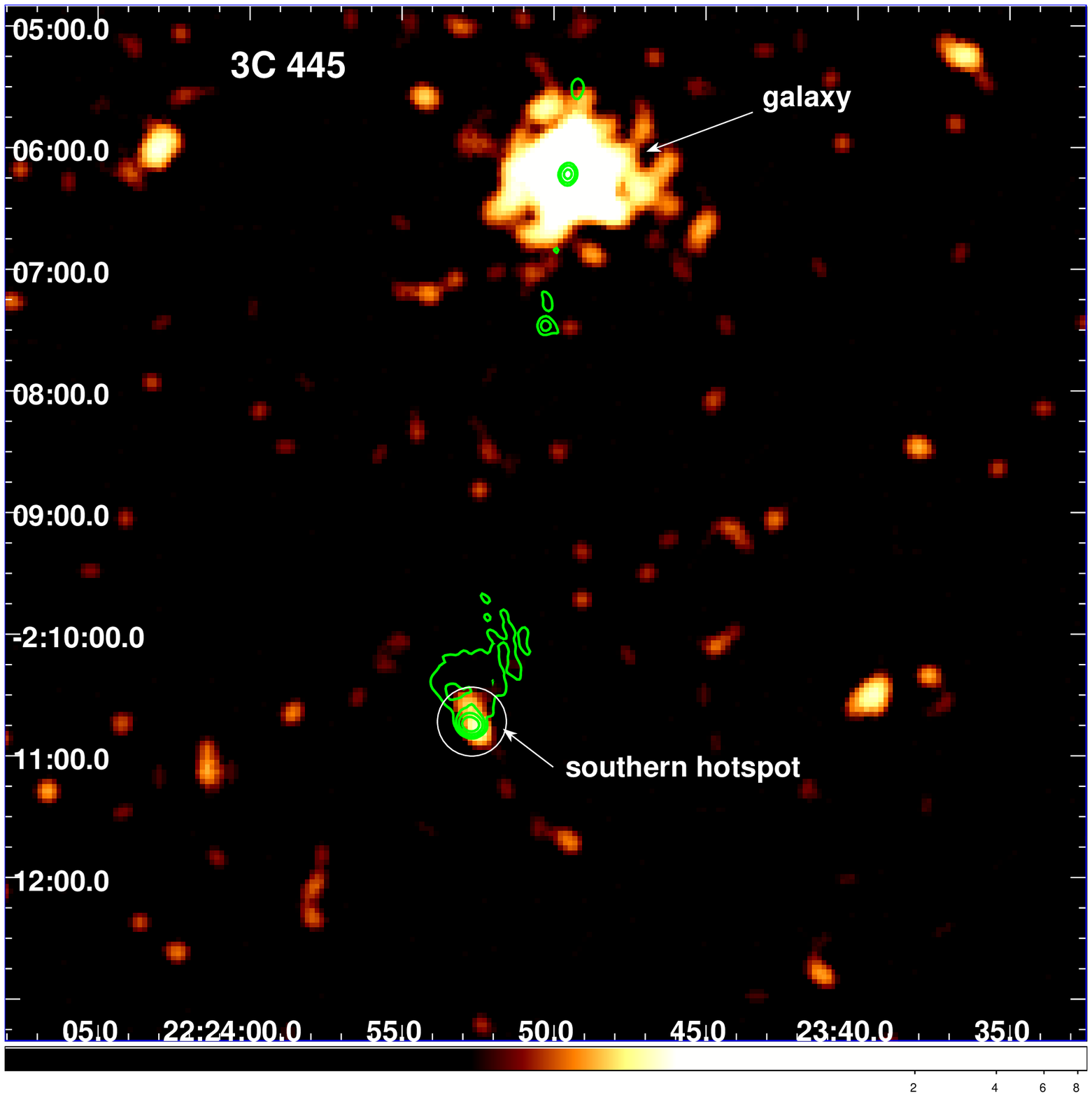}
\vspace{14cm}
\caption{The {\it Swift} images (0.3-10 keV) of 3C\,105 and
  3C\,445 are superimposed to the radio contours at 8.4 GHz, for
3C\,105 ({\it top}), and at 1.4 GHz for 3C\,445 ({\it bottom}). First
contour level is 3.7 mJy/beam and 3.2 mJy/beam for 3C\,105 and 3C\,445
respectively. Contour levels increase by a factor of 3.}
\label{appendice}
\end{center}
\end{figure}

\noindent 3C\,105 has been observed by {\it Swift} in four occasions 
(Obs. ID 00035625001-2-3-4)
for a total exposure of $\sim$ 22 ks while 3C\,445 
only for $\sim$ 12 ks (Obs. ID 00030944001-2).
During all these observations, the {\it Swift} satellite was
operated with all the instruments in data taking mode. We
consider only XRT \citep{burrows05} data, since our
sources were not bright enough to be detected by the BAT
high energy experiment. In particular, {\it Swift}-XRT
observations have been performed in photon-counting mode (PC).\\
The XRT data analysis has been performed with the XRT-
DAS software, developed at the ASI Science Data
Center (ASDC) and distributed within the HEAsoft pack-
age (v. 6.9). Event files were calibrated and cleaned with
standard filtering criteria using the xrtpipeline task, combined 
with the latest calibration files available in the {\it Swift}
CALDB distributed by HEASARC. Events in the energy
range 0.3-10 keV with grades 0-12 (PC mode) were used in
the analysis (see Hill et al. 2004 for more details). No signatures 
of pile-up were found in our {\it Swift} XRT observations.
Events are extracted using a 17 arcsec radius circle centered on
the radio position of the southern hotspots in both cases of 3C\,105
and 
3C\,445 (see Fig. \ref{appendice}).
we measured 15 counts in the southern hotspot of 3C\,105 and 
12 counts for that of 3C\,445,
while the background estimated from a nearby source-free circular 
region of the same radius is 1.8 counts and 0.9 respectively. 

\end{document}